\documentclass[10pt,conference]{IEEEtran}
\IEEEoverridecommandlockouts
\usepackage{cite}
\usepackage{amsmath,amssymb,amsfonts}
\usepackage{graphicx}
\usepackage{textcomp}
\usepackage{xcolor}

\usepackage{algorithm}
\usepackage{algpseudocode}
\usepackage{cite}
\usepackage{bm}
\usepackage{bbm}
\usepackage{grffile}
\usepackage{booktabs}
\usepackage[hyphens]{url}
\usepackage{subcaption} % subfigure
\usepackage[acronym]{glossaries}
\usepackage{mathtools}
\usepackage{multirow}
\usepackage{booktabs} % table hlines
\usepackage{xspace}

\graphicspath{{images/}}
\DeclareGraphicsExtensions{.pdf,.jpg,.png}

\newcommand{\fsAcc}{0.23\textwidth}

\renewcommand{\[}{\left[}
\renewcommand{\]}{\right]}
\renewcommand{\(}{\left(}
\renewcommand{\)}{\right)}

\newcommand{\lv}{\left\lvert}
\newcommand{\rv}{\right\rvert}

\DeclareMathOperator*{\minimize}{minimize}

\newcommand{\Real}{\mathbb{R}}

\newcommand{\Inputs}{\mathcal{X}}
\newcommand{\Outputs}{\mathcal{Y}}
\newcommand{\inputarg}{\bm{x}}
\newcommand{\predy}[2]{\bm{y}_{#1}^{#2}}
\newcommand{\LSpc}[1][\mu]{L^2_{#1}}
\newcommand{\Model}{f}
\newcommand{\pdf}{\mu}
\newcommand{\dist}{d}
\newcommand{\distF}{d_{\LSpc[\pdf]}}
\newcommand{\costfunc}{c}

\newcommand{\Devices}{\mathcal{U}}
\newcommand{\NumDevices}{n}
\newcommand{\Neigh}[1]{\mathcal{N}_{#1}}
\newcommand{\NumNeigh}{n}
\newcommand{\DimX}{X}
\newcommand{\DimY}{Y}
\newcommand{\sharedD}{\mathcal{D}_\mathrm{s}}
\newcommand{\Dsi}[1]{\mathcal{D}_{#1}}
\newcommand{\Dsm}{\hat{\mathcal{D}}}
\newcommand{\BatchShared}{\tilde{\mathcal{D}}^t}
\newcommand{\NumBatchShared}{n_\mathrm{d}^t}
\newcommand{\NumDataTicks}{d}
\newcommand{\NumRound}{T}
\newcommand{\TimeStep}{\tau}

\newcommand{\Lmu}{\mathsf{L}_{\mu}}
\newcommand{\Lmui}{\mathsf{L}_{\mu_i}}

\newcommand{\lr}{\eta}
\newcommand{\lrc}{\varepsilon}

\newcommand{\ws}[1][i]{{\hat{\bm{w}}_{#1}}}
\newcommand{\prmw}{\bm{w}}
\newacronym{IID}{IID}{independent and identically distributed}
\newacronym[plural={IoT}]{IoT}{IoT}{Internet of Things}
\newacronym{FL}{FL}{federated learning}
\newacronym{FD}{FD}{federated distillation}
\newacronym{DFL}{DFL}{decentralized \gls{FL}}
\newacronym{CFL}{CFL}{centralized \gls{FL}}
\newacronym{DFD}{DFD}{decentralized \gls{FD}}
\newacronym{DCCR}{DCCR}{dynamic communication cost reduction method}
\newacronym{DL}{DL}{deep learning}
\newacronym{NN}{NN}{neural network}
\newacronym{DNN}{DNN}{deep neural network}
\newacronym{ML}{ML}{machine learning}
\newacronym{KD}{KD}{knowledge distillation}
\newacronym{MSE}{MSE}{mean squared error}
\newacronym{KL}{KL}{Kullback-Leibler}
\newacronym{WSN}{WSN}{wireless sensor networks}
\newacronym{P2P}{P2P}{peer-to-peer}
\newacronym{SGD}{SGD}{stochastic gradient descent}
\newacronym{FMNIST}{F-MNIST}{fashion MNIST}
\newacronym{GAN}{GAN}{generative adversarial network}
\newacronym{RHS}{RHS}{right-hand side}
\newacronym{UMAP}{UMAP}{uniform manifold approximation and projection}
\newacronym{CMFD}{CMFD}{consensus-based multi-hop federated distillation}
\newacronym{SVD}{SVD}{singular value decomposition}
\newacronym{PCA}{PCA}{principal component analysis}
\newacronym{tSNE}{t-SNE}{t-distributed stochastic neighbor embedding}
\newacronym{PRNG}{PRNG}{pseudo-random number generator}

\newcommand{\etal}{\textit{et al.\ }}

\def\BibTeX{{\rm B\kern-.05em{\sc i\kern-.025em b}\kern-.08em
    T\kern-.1667em\lower.7ex\hbox{E}\kern-.125emX}}
 
\begin{document}

\title{Convergence Visualizer\\of Decentralized Federated Distillation\\with Reduced Communication Costs
\thanks{This work was partly supported by NICT, Japan (01101) and JSPS KAKENHI Grant Number JP21K17734.}
}
% \thanks{This work was supported by JSPS KAKENHI Grant Numbers JP21K17734,JP20H00622.}

\author{\IEEEauthorblockN{1\textsuperscript{st} Akihito Taya}
\IEEEauthorblockA{\textit{Institute of Industorial Science,} \\
\textit{The University of Tokyo}\\
Tokyo, JAPAN \\
taya-a@iis.u-tokyo.ac.jp
}
\and
\IEEEauthorblockN{2\textsuperscript{nd} Yuuki Nishiyama}
\IEEEauthorblockA{\textit{Center for Spatial Information Science,} \\
\textit{The University of Tokyo}\\
Chiba, JAPAN \\
yuukin@iis.u-tokyo.ac.jp
}
\and
\IEEEauthorblockN{3\textsuperscript{rd} Kaoru Sezaki}
\IEEEauthorblockA{\textit{Center for Spatial Information Science,} \\
\textit{The University of Tokyo}\\
Chiba, JAPAN \\
sezaki@iis.u-tokyo.ac.jp}
}

\maketitle

\begin{abstract}
\Gls{FL} achieves collaborative learning without the need for data sharing, thus preventing privacy leakage.
To extend \gls{FL} into a fully decentralized algorithm, researchers have applied distributed optimization algorithms to \gls{FL}
by considering \gls{ML} tasks as parameter optimization problems.
Conversely, the \gls{CMFD} proposed in the authors' previous work
makes \gls{NN} models get close with others in a function space rather than in a parameter space.
Hence, this study solves two unresolved challenges of \gls{CMFD}:
(1) communication cost reduction and (2) visualization of model convergence.
Based on a proposed \gls{DCCR}, the amount of data transferred in a network is reduced; however, with a slight degradation in the prediction accuracy.
In addition, a technique for visualizing the distance between the \gls{NN} models in a function space is also proposed.
The technique applies a dimensionality reduction technique by approximating infinite-dimensional functions as numerical vectors
to visualize the trajectory of how the models change by the distributed learning algorithm.
\end{abstract}

\begin{IEEEkeywords}
federated learning, multi-hop networks, consensus algorithm, decentralized machine learning, visualization
\end{IEEEkeywords}

\glsresetall

% \mytodo{1. Explain difference from original CMFD}

% \mytodo{2. Explain details of numerical results (whole part?)}

% \mytodo{3. (5): missing the set of subscripts i???, (May be explanation of $\BatchShared$ is required.)}

% \mytodo{4. Sec.6 Explain "conventional methods" means CMFD}

% \mytodo{5. Fig. 2,3: explanation of difference between F-MNIST and CIFAR10}

% \mytodo{6. Fig. 4,5: explanation of why 1,2,6 did not converge}

% \mytodo{7. Fig 6,7: add explanation}

\section{Introduction}
Owing to its communication-efficiency and privacy-preserving properties,
\gls{FL} is a promising framework for the collaborative learning of \gls{IoT} sensor devices, wearable sensors, smartphones, smart homes, and connected vehicles~\cite{mcmahan2016communication,kairouz2019advances,uddin2019iot}.
McMahan \etal~\cite{mcmahan2016communication} originally proposed \gls{FL} as a distributed \gls{ML} algorithm to prevent the leakage of user privacy included in local data samples
by avoiding the upload of user data to a central server.
The emergence of \gls{FL} has garnered considerable attention,
and many researchers have extended \gls{FL} to address the remaining challenges, e.g., non-\gls{IID} datasets, as well as further reduction of communication costs.

One problem is that typical \gls{FL} algorithms require a central server
to manage the learning process and aggregate the collected parameters to update a global \gls{NN} model.
In \gls{CFL}, the server capacity for computation and communication can be a bottleneck,
and there is a concern regarding the Big Tech monopoly of well-trained models.
However, \gls{DFL}~\cite{beltran2022decentralized,savazzi2020federated,elgabli2020gadmm} is an important extension for solving such problems.
In \Gls{DFL}, client devices communicate directly with others to share the model-update information without a central server.
Consequently, \gls{DFL} is suitable for \gls{IoT} applications in which sensor devices are connected via \gls{WSN}.

We proposed \gls{CMFD}~\cite{taya2022decentralized} as a \gls{DFL} using \gls{KD}.
% An advantage of distillation is a \mytodo{advantage}.
The concept of \gls{CMFD} is to solve convex functional optimization rather than non-convex parameter optimization.
Although \glspl{NN} are non-convex, \gls{ML} tasks can be convex problems in terms of functional optimization.
Therefore, \gls{CMFD} attempts to solve the optimization problem in a distributed manner by directly aggregating functions rather than the \gls{NN} parameters.
This concept can be realized by decentralizing algorithms of \gls{FD}.
\Gls{FD} is a variation of \gls{FL}, in which devices share the output values of their local models, called distilled values, rather than the parameters.

Although \gls{FD} requires lower communication costs than parameter-aggregation-based \gls{FL},
there is a potential to reduce the amount of transferred data.
In addition, the communication cost can be reduced if the shared distilled values are limited;
however, the prediction performance is degraded.
To overcome the accuracy-communication tradeoff,
% \begin{revhl}{1}
this paper proposes a \gls{DCCR} as an extension of \gls{CMFD}.
% \end{revhl}
\Gls{DCCR} enables devices to share a few distilled values with others at the beginning of training
and gradually increase the amount of shared data to improve prediction accuracy.

A key feature of the \gls{DCCR} is communication-less synchronization of the distributed random selection.
\Gls{DFD} must share input values with other devices to aggregate the distilled values.
Hence, the devices should synchronize the information indicating which samples are used to calculate the distilled values.
To suppress bothersome communication,
the \gls{DCCR} leverages a \gls{PRNG} to obtain randomly selected synchronized data samples in a distributed manner.

Another challenge in \gls{DFD} is validating model convergence properties.
In contrast to \gls{CFL}, where devices easily synchronize their models to the global one that is managed by a server,
local models may not converge to the same one in finite time with a decentralized algorithm,
even though the theoretical analysis guarantees convergence assuming infinite time.
Moreover, when \gls{KD} is applied, our interest lies in the convergence of prediction functions rather than that of the \gls{NN} parameters.
Because the convergence of models cannot be quantified directly,
a novel framework for validating their convergence should be developed.

To validate the convergence properties of the trained models,
we develop a visualization framework that visualizes how the prediction models trained by the local devices approach others in a function space.
Although dimensionality-reduction techniques have been widely adopted for high-dimensional data visualization,
their targets are high-dimensional data,
which implies that they cannot be applied to \gls{NN} models because they have infinite dimensions as functions.
Therefore, we approximate these functions as finite-dimensional vectors such that the distances between the two functions remain unchanged.
Subsequently, a dimensionality-reduction technique is applied to the finite-dimensional vectors to obtain a low-dimensional projection.

The contributions of this paper are summarized as follows:
\begin{itemize}
\item We propose a \gls{DCCR} as an extention of \gls{CMFD}.
      Devices first utilize a few values for distillation to suppress the communication costs
      and gradually increase the amount of shared information to improve the prediction accuracy toward the end of the training.
      In addition, we leverage a \gls{PRNG} to share information of random sampling with zero-cost communication.
\item We provide a scheme for visualizing how local models approach each other in a function space.
      This scheme leverages a dimensionality-reduction technique to project local models onto a 2D/3D space,
      demonstrating that the local models get close to each other in the function space, even with the proposed communication-efficient method.
\end{itemize}

\section{Related works}
\subsection{Decentralized and distillation-based federated learning}
In the context of the \gls{IoT}, where sensor devices are connected via \gls{WSN},
\gls{DFL} has advantages over centralized algorithms.
Although \gls{CFL} is the predominant approach in \gls{FL},
the communication and computational capacity of a central server can become challenging when the number of participating devices increases.
Hence, \gls{DFL} has been developed to address this limitation~\cite{beltran2022decentralized,savazzi2020federated,lalitha2019peer,niwa2020edge}.
The \gls{DFL} assumes that the client devices are connected via multi-hop networks and communicate directly with each other to share information.

By extending FedAvg~\cite{mcmahan2016communication},
\gls{DFL} algorithms that share the parameters of the \gls{NN} models among devices have been proposed~\cite{savazzi2020federated,lalitha2019peer,niwa2020edge}.
However, these schemes require numerous communication resources because of their numerous parameters.
Conversely, \gls{DFD} algorithms send the output values of local prediction models, referred to as distilled values, to adjacent devices~\cite{anil2018large,zhang2018deep}.
Our previous study~\cite{taya2022decentralized} was also categorized as \gls{DFD}.
\Gls{DFD} reduces the communication cost of \gls{FL} because the number of distilled values is typically less than the number of parameters of the \gls{DNN};
however, \gls{DFD} requires a number of distilled values to improve the training performance.
Therefore, a communication-efficient method for the \glspl{DFD} should be developed.

There are several methods for reducing the communication costs of \gls{FL}~\cite{jiang2022model,zhang2022fedduap,xu2021accelerating,nishio2019client}.
For instance, model pruning~\cite{jiang2022model,zhang2022fedduap} removes redundant structures and parameters from \gls{NN} models.
Although this technique reduces the communication cost of sharing model gradients,
it is not suitable for \gls{FD} where model gradients are not shared.
Another approach to suppress network traffic is client selection, in which only a few clients upload their model updates to the server~\cite{xu2021accelerating,nishio2019client}.
However, this cannot be applied to decentralized algorithms.
In addition, quantization is a basic approach for reducing communication costs.
% Model and gradient quantization for \gls{FL} were proposed in~\cite{}.
In \gls{FD}, distilled values, rather than model parameters, are quantized~\cite{sattler2020communication}.
Hence, we focus on reducing the number of distilled values utilized in each communication round,
which can be easily combined with compression methods.

\subsection{Visualization of neural networks}
Uddin \etal~\cite{uddin2020mutual} visualized the progress of \gls{CFL} by plotting the mutual information
between the ground truth information and output logits produced by every global model in each communication round.
Although this technique can visualize the relationship between the ground truth and the global model at the central server,
the relationship between more than two devices cannot be visualized in decentralized scenarios.

% [] visualizes the results of \gls{FL} in the following approach.
% They select three labels to learn from, and plot their probability outputs as coordinates on a plane in a three-dimensional space.

% \subsection{Dimensionality reduction}
Dimensionality-reduction techniques are suitable approaches for visualization.
They have been developed to project high-dimensional data onto a lower-dimensional space
while preserving the original data structure and relationships within the data.
\Gls{PCA} is a traditional technique that utilizes \gls{SVD}~\cite{bishop2006pattern} to linearly transform high-dimensional data into a lower-dimensional space.
In addition, \gls{UMAP}~\cite{mcinnes2018umap} and \gls{tSNE}~\cite{maaten2008visualizing} are nonlinear dimensionality-reduction techniques used for high-dimensional-data visualization.
% \gls{tSNE}~\cite{maaten2008visualizing} works by converting pairwise similarities in the high-dimensional space into probabilities,
% and then minimizing the divergence between these probabilities and those computed in the lower-dimensional space using a t-distribution-based cost function, resulting in a more faithful representation of the original data structure.
\Gls{UMAP} approximates the high-dimensional manifold structure and optimizes lower-dimensional embedding by minimizing the cross-entropy between the high- and low-dimensional representations.
We adopt \gls{UMAP} because it tends to output more separated clusters than \gls{tSNE}.

\section{Problem statement and the concept of CMFD}\label{sec:problem}

We consider a system where some devices with sensors are connected via a multi-hop network.
Each device $i$ trains its local model $\Model_i$ utilizing a local dataset $\Dsi{i}$.
To avoid privacy leakage, the dataset $\Dsi{i}$ is not shared with other devices
In contrast to a typical \gls{DFL}, where the devices share information of parameter updates,
the devices share distilled values, i.e., the outputs of $\Model_i$.
When calculating the distilled values, all devices utilize the same input $\sharedD$,
which is assumed to be a public dataset or an artificially generated dataset~\cite{jeong2018communication}.

The key concept proposed in our previous work~\cite{taya2022decentralized} was
to operate a consensus-based optimization algorithm in a function space to avoid non-convexity problems.
Although training \glsplural{NN} is typically defined as the optimization of parameters,
parameter optimization encounters challenges because of non-convexity in the \gls{DFL} context.
Therefore, prediction functions rather than parameters should be optimized as follows:
\begin{align}
  \minimize_{f\in\LSpc} \quad \Lmu(\Model) = \sum_{i\in\Devices} \Lmui(\Model), \label{eq:opt_functions}
\end{align}
where $\pdf$, $\pdf_i$, and $\LSpc[\pdf]$ represent the global and local probability measure of the input space, and the $L^2$ space with respect to the probability measure $\pdf$, respectively.
We denote $\Lmu$ and $\Lmui$ as the global and local loss functions, respectively.
Notably, typical loss functions (e.g., \gls{KL} divergence~\cite{van2014renyi} and \gls{MSE}) satisfy the following inequities for all $\Model_1, \Model_2$, and $k\in[0,1]$:
\begin{align}
  \Lmu(tf_1 + (1-t)f_2) &\le t \Lmu(f_1) + (1-t) \Lmu(f_2), \\
  \Lmui(tf_1 + (1-t)f_2) &\le t \Lmui(f_1) + (1-t) \Lmui(f_2).
\end{align}
This implies that (\ref{eq:opt_functions}) is a convex functional optimization problem,
although optimizing the parameters $\prmw_i$ of the \gls{NN} is a non-convex problem.
Hence, \gls{CMFD} focuses on updating the prediction models $\Model$ directly in the direction of reducing $\Lmu(\Model)$ when aggregating the local update information.

In \gls{CMFD}, the devices are allowed to have different parameter sets at the end provided that the trained models are the same
because functions are treated as identical in $\LSpc[\mu]$ iff the distance between the functions is equal to zero as follows:
\begin{align}
  \distF(\Model_{1},\Model_{2})
    \coloneqq \sqrt{\int_\Inputs \lv\dist_\Outputs(\Model_{1}(\inputarg), \Model_{2}(\inputarg))\rv^2 \mathrm{d}\pdf}
    = 0, \label{eq:truedist}
\end{align}
where $\Inputs\subseteq\Real^\DimX$, $\Outputs\subseteq\Real^\DimY$, $\distF$ and $\dist_\Outputs$ denote
the input and output space, and the distance of functions and output values, respectively.
It should be noted that functions with different parameters can be the same because of their non-convexity.

In \gls{CMFD}, the devices iterate the following two steps in the $t$th communication round:
\begin{enumerate}
  \item Update the local \gls{NN} parameters $\prmw_i^{t}$ by \gls{SGD} utilizing $\Dsi{i}$.
  \item Update $\prmw_i^{t}$ by \gls{KD} using the received distilled values $\predy{j,\inputarg}{t}$ from adjacent devices $j\in\Neigh{i}$.
\end{enumerate}
These steps emulate consensus-based optimization in the function space $\LSpc$,
and therefore the \gls{NN} models $\Model_i$ are optimized appropriately because of the convexity of (\ref{eq:opt_functions}).% in the function space.

\section{Dynamic Communication Cost Reduction}\label{sec:dccr}
Compared with \gls{FL}, \gls{FD} reduces communication costs
because the dimensions of the distilled values are fewer than the parameters of \gls{DNN}.
However, there are demands to further reduce the amount of data transferred in \gls{WSN} because of limited communication resources.
Communication costs can be reduced by limiting the dataset $\sharedD$ utilized to calculate the distilled values $\predy{i,\inputarg(\in\sharedD)}{t}$,
which, unfortunately, results in performance degradation.
To overcome the tradeoff between communication costs and prediction accuracy,
% \begin{revhl}{1}
we extend \gls{CMFD} by utilizing \gls{DCCR}
% \end{revhl}
in which the devices limit the distilled values to be shared at the beginning of training to reduce communication costs,
whereas the number of the distilled values used for training is gradually increased to improve the prediction accuracy.

\begin{figure}[!t]
  \begin{algorithm}[H]
    \caption{Pseudo code of \gls{DCCR}}\label{alg:dccr}
    \begin{algorithmic}[1]
      \Require %Prediction function with \gls{NN} parameters $\bm{w}$: $f(\cdot;\bm{w})$,

               % Initial parameters: $w_i^1$,

               % local loss function: $l(\bm{y},\bm{y}')$

               Local and shared train data: $\Dsi{i}$, $\sharedD$,

               Learning and sharing rate: $\lr_t, \lrc$
      \While{not converged}
        \ForAll{device $i=1,\ldots,\NumDevices$}
          \For {minibatch $\Dsm$ in $\Dsi{i}$} \label{algln:local_sgd_bgn}
            % \State $\ws^{t} \leftarrow \bm{w}_i^t - \lr_t \nabla_{\bm{w}} \displaystyle{\sum_{(\bm{x},\bm{y})\in\Dsm}}l(f(\bm{x};\bm{w}_i^t),\bm{y})$~\label{algln:local_sgd_grad}
            \State $\ws^{t} \leftarrow \bm{w}_i^t - \lr_t \nabla_{\bm{w}} \mathop{Loss}(f(\cdot;\bm{w}_i^t), \Dsm)$~\label{algln:local_sgd_grad}
          \EndFor \label{algln:local_sgd_end}
          \State Select subset $\BatchShared$ of $\sharedD$ using the shared \gls{PRNG}.~\label{algln:select_batch}
          \ForAll {$x\in\BatchShared$}~\label{algln:pub_calc_bgn}
            \State $\predy{i,\inputarg}{t} \leftarrow f(\bm{x};\ws^{t})$
          \EndFor \label{algln:pub_calc_end}
          \State send $\predy{i,\inputarg}{t}$ to neighbor devices \label{algln:send_distil}
        \EndFor
        \ForAll{device $i=1,\ldots,\NumDevices$}
          \State receive $\predy{j,\inputarg}{t}$ from neighbor devices
          \For {minibatch $\Dsm$ in $\BatchShared$}
          % \State $\displaystyle{c(\ws) \coloneqq \sum_{(\bm{x},\bm{y})\in\Dsm}\lv f(\bm{x};\ws^t) - \frac{1}{\NumNeigh_i}\sum_{j\in\Neigh{i}}\predy{j,\bm{x}}^t\rv^2}$ \label{algln:distil_bgn}
            \State $\bm{w}_i^{t+1} \leftarrow \ws^{t} - \lrc \NumNeigh_i \nabla_{\bm{w}} \costfunc(\ws^{t})$ \label{algln:distil_end} 
          \EndFor
        \EndFor
      \EndWhile
    \end{algorithmic}
  \end{algorithm}
\end{figure}

The pseudo-code for the proposed algorithm is presented in Alg.~\ref{alg:dccr}.
Each device performs local \gls{SGD} to update the local parameters $\prmw_i^t$ on lines~\ref{algln:local_sgd_bgn}--\ref{algln:local_sgd_end}.
After the local \gls{SGD} phase, the device randomly selects $\NumBatchShared$ data samples from $\sharedD$ to make a temporary dataset $\BatchShared$ on Line~\ref{algln:select_batch}.
% \begin{revhl}{3}
$\BatchShared$ is selected such that the same samples are included across devices in a distributed manner, as explained in the last part of this section.
% \end{revhl}
Using the selected samples $\BatchShared$,
each device calculates the output values of its local model $\Model_{i}\coloneqq\Model(\cdot; \ws^t)$ in lines~\ref{algln:pub_calc_bgn}--\ref{algln:pub_calc_end}.
The output values, which are referred to as the distilled values $\predy{i,\inputarg}{t}$, are sent to adjacent devices (line~\ref{algln:send_distil}).
After receiving the distilled values, each device updates its local model (line~\ref{algln:distil_end}).

\begin{figure}[!t]
\centering
\includegraphics[width=0.40\textwidth]{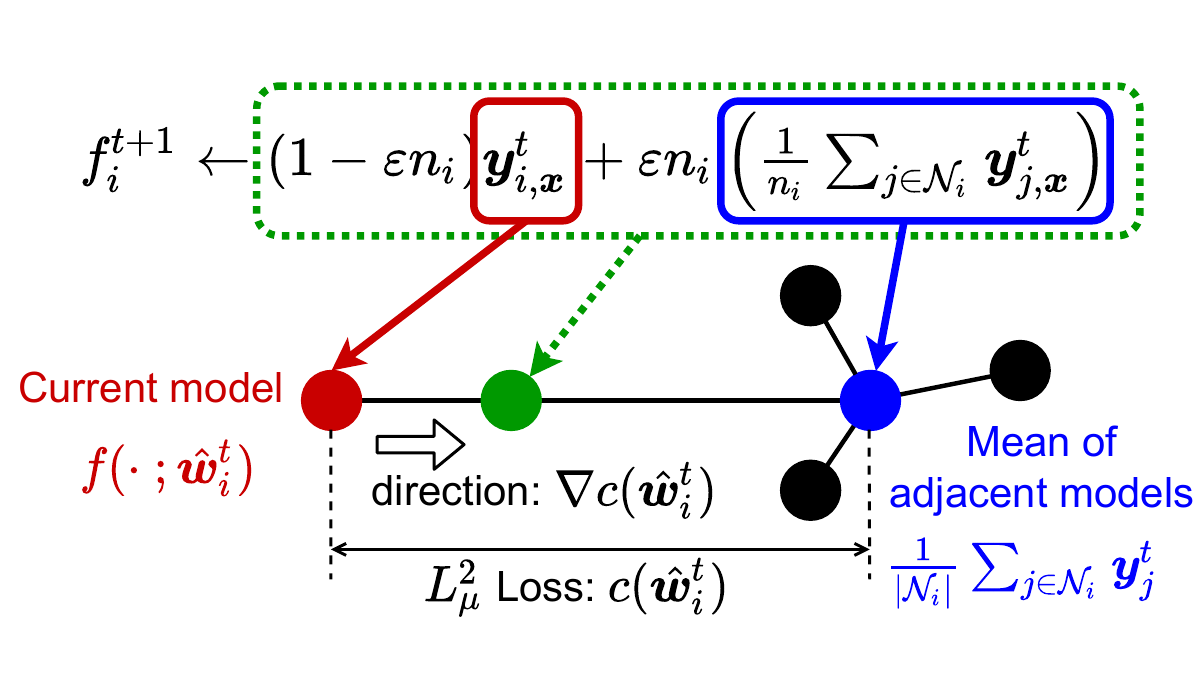}
\caption{Concept of distillation-based function aggregation.
The local model approaches the other models in a function space by \gls{SGD}, treating the mean of the adjacent output values as a target value.}\label{fig:model_aggregation}
\end{figure}

In contrast to parameter-aggregation-based \gls{FL} algorithms, which calculate the weighted average of the parameters $\ws^t$,
\gls{FD} performs \gls{SGD} for model aggregation to make the models closer in a function space.
The concept of model aggregation is illustrated in Fig.~\ref{fig:model_aggregation}.
The devices share the distilled values $\predy{i,\inputarg}{t}$,
and each device calculates the average of the received distilled values, which becomes the new target value.
To make the local models closer to other models in the function space,
the devices update their local models using the gradient of the distance between their local models and the obtained target values,
that is, the devices perform \gls{SGD} according to the following criteria:
\begin{align}
  \costfunc(\ws^{t}) \coloneqq \sum_{\inputarg\in\BatchShared} \bigg\| f(\inputarg;\ws^{t}) - \frac{1}{\lv\Neigh{i}\rv}\sum_{j\in\Neigh{i}}\predy{j,\inputarg}{t}\bigg\|^2, \label{eq:distillation_loss}
\end{align}
where $\lv\mathcal{\cdot}\rv$ denotes the cardinality of a set.
The second term in the norm is the target value in the $t$th communication round.

When selecting $\BatchShared$ from $\sharedD$, the \gls{DCCR} dynamically changes the number of data samples $\NumBatchShared$,
which is determined as follows:
\begin{align}
  % \NumBatchShared \coloneqq \NumDataTicks \(\left\lfloor \frac{t \lv\sharedD\rv}{\NumRound \NumDataTicks} \right\rfloor + 1\),
  \NumBatchShared \coloneqq \NumDataTicks \(\left\lfloor \frac{t}{\TimeStep} \right\rfloor + 1\), \label{eq:num_increase_rate}
\end{align}
where $\lfloor\cdot\rfloor$ denotes the floor function.
Coefficients $\NumDataTicks$ and $\TimeStep$ control parameters for determining the rate of increase of $\NumBatchShared$.
When $\TimeStep$ is equal to $\NumRound\NumDataTicks/\lv\sharedD\rv$, all data samples in $\sharedD$ are utilized in the last $\tau$ round of training.
Because $\NumBatchShared$ increases with communication round $t$,
\gls{DCCR} suppresses the traffic load at the beginning of training
while updating the \gls{NN} models carefully utilizing sufficient distilled values at the end.

\noindent\textbf{Zero-cost information sharing of random samples:}

Selecting $\BatchShared$ in a distributed manner requires a strategy to share the selected samples
to address the accuracy-communication tradeoff.
When all devices utilize all the samples, the sample-selection information does not need to be shared.
In \gls{DCCR}, however, if each device selects different data samples in a distributed manner,
the cost function $\costfunc(\ws^{t})$ cannot be calculated.
To suppress the costs of sharing sample-selection information, the \gls{DCCR} leverages a \gls{PRNG}
that is included in the standard libraries of daily used programming languages.

Although the \gls{PRNG} approximates the properties of random sequences, we utilize its deterministic features in \gls{DCCR}.
The \gls{PRNG} generates the same series from the same initial key, which is called the seed.
Therefore, if the devices share the information of the initial key of the \gls{PRNG} before beginning training,
the devices can obtain the same randomly selected data samples $\BatchShared$ in a distributed manner without communicating with the others at each iteration.

\section{Visualization of function convergence} \label{sec:visualization}

As described in Sec.~\ref{sec:problem}, 
our algorithm focuses on updating the \gls{NN} models to converge in the function space rather than in the parameter space.
This section provides a scheme that maps functions from $\LSpc$ to positions in 2D or 3D space
to visualize the proximity of the \gls{NN} models to others in the function space.
Using this scheme, we can observe the dynamics of the model updates in the training phase.

Dimensionality-reduction techniques are typically adopted to project high-dimensional data onto a low-dimensional space.
However, the prediction models $\Model_i$ have infinite dimensions, indicating that dimensionality-reduction techniques cannot be applied directly.
Considering that we are interested in the distance between the models in the convergence analysis,
it is sufficient that a visualized image reflects the distance between the prediction models.
Therefore, the proposed visualization scheme evaluates the empirically approximated distances of the prediction models.

% Distance $\distF$ between two functions is defined as follows:
% \begin{align}
%   \distF(\Model_{1},\Model_{2})
%   \coloneqq \sqrt{\int_\Inputs \lv\dist_\Outputs(\Model_{1}(\inputarg), \Model_{2}(\inputarg))\rv^2 \mathrm{d}\pdf}, \label{eq:truedist}
% \end{align}
% where $\dist_\Outputs$ the distance of output values.
% Note that the distance can be calculated without the knowledge of the implementation.
The ideal distance between the two models is defined by (\ref{eq:truedist}),
which cannot be calculated when using a \gls{NN} with numerous parameters.
Hence, we use the following empirical representation:
\begin{align}
  \distF(\Model_{i},\Model_{j})
  \approx \sqrt{\sum_{\inputarg\sim\pdf} \lv\dist_\Outputs(\Model_{i}(\inputarg), \Model_{j}(\inputarg))\rv^2}. \label{eq:approxdist}
\end{align}
When $\Outputs$ is a subset of $\Real^{\DimY}$, the \gls{RHS} of (\ref{eq:approxdist}) is equivalent to the distance between the flattened output vectors corresponding to specific input values.
In the context of \gls{FD}, the shared dataset $\sharedD$ can be used as the shared input values.
Whereas the distributions $\pdf_i$ of local datasets $\Dsi{i}$ differ among devices under a non-\gls{IID} condition,
$\sharedD$ is expected to have an \gls{IID} by sampling appropriately from the public dataset.

Let $\predy{i,\sharedD}{t}$ denote the vectorized output values of device $i$ defined as follows:
\begin{align}
  \predy{i,\sharedD}{t} \coloneqq \[ \Model_i^t(\inputarg_1)^\top \dots \Model_i^t(\inputarg_{\lv\sharedD\rv})^\top \]^\top \in \Real^{\DimY \lv\sharedD\rv}, \label{eq:vecform}
\end{align}
where $A^\top$ denotes the transpose of matrix $A$.
As mentioned previously, the output vectors $\predy{i,\sharedD}{t}$ and $\predy{j,\sharedD}{t}$ can be utilized rather than the models $\Model_i$ and $\Model_j$
when projecting models onto a low-dimensional space with dimensionality-reduction techniques
because the distance of the output vector coincides with the empirical approximation of the models (\ref{eq:approxdist}).
\Gls{UMAP}~\cite{mcinnes2018umap} is a distance-based dimension reduction tool
that can be utilized to visualize high-dimensional data.
By treating (\ref{eq:vecform}) as a feature vector, its dimensions can be reduced to two or three using the \gls{UMAP}.
Consequently, positions of the dimension-reduced vectors can then be plotted in a 2D or 3D space.

% When the functions are reduced into three dimension data, we can also visualization them by three values 

% \begin{figure}[!t]
% \centering
% \includegraphics[width=0.12\textwidth]{images/topologies/r3.pdf}
% \caption{Ring lattice topology of the evaluated network. Each device communicates with six adjacent devices.}\label{fig:topo_r3}
% \end{figure}

\section{Evaluation}
\subsection{Prediction accuracy with \gls{DCCR}}
\begin{table}[!b]
  \caption{Simulation parameters}
  \centering
  \begin{tabular}{cc}
    \toprule
    \textbf{Parameters} & \textbf{Values} \\
    \midrule
    Num.\ devices & 10 \\
    Network topology & Ring lattice \\
    \multirow[vpos]{2}{*}{Num.\ training data per device $\lv\Dsi{i}\rv$} & 1000 (F-MNIST) \\
                                                                          & 4000 (CIFAR10) \\
    \multirow[vpos]{2}{*}{Num.\ share data $\lv\sharedD\rv$} & 1000 (F-MNIST) \\
                                                             & 10000 (CIFAR10) \\
    \bottomrule
  \end{tabular}\label{tbl:simparms}
\end{table}

In this section, we discuss the evaluation results of the \gls{DCCR} using \gls{FMNIST}~\cite{xiao2017fashion} and CIFAR10~\cite{krizhevsky2009learning}.
The simulation parameters are listed in Table~\ref{tbl:simparms}.
% We evaluated a ring lattice topology shown in Fig.~\ref{fig:topo_r3}.
We evaluated the ring lattice topology in which each device communicated with six adjacent devices.
Device $i$ possesses a non-\gls{IID} local dataset including two categories labeled $i$ and $((i+1) \bmod 10)$,
assuming a situation where nearby devices have similar data distributions,
where $\bmod$ denotes the modulo operator.
% \begin{revhl}{2}
We utilized the optimization software Optuna~\cite{Akiba2019-jz} to optimize the learning rate $\lr$ and sharing rate $\lrc$
that appear in Alg.~\ref{alg:dccr}.
% \end{revhl}

\begin{figure}[!t]
\centering
\begin{subfigure}[t]{\fsAcc}
  \centering
  \includegraphics[width=\textwidth]{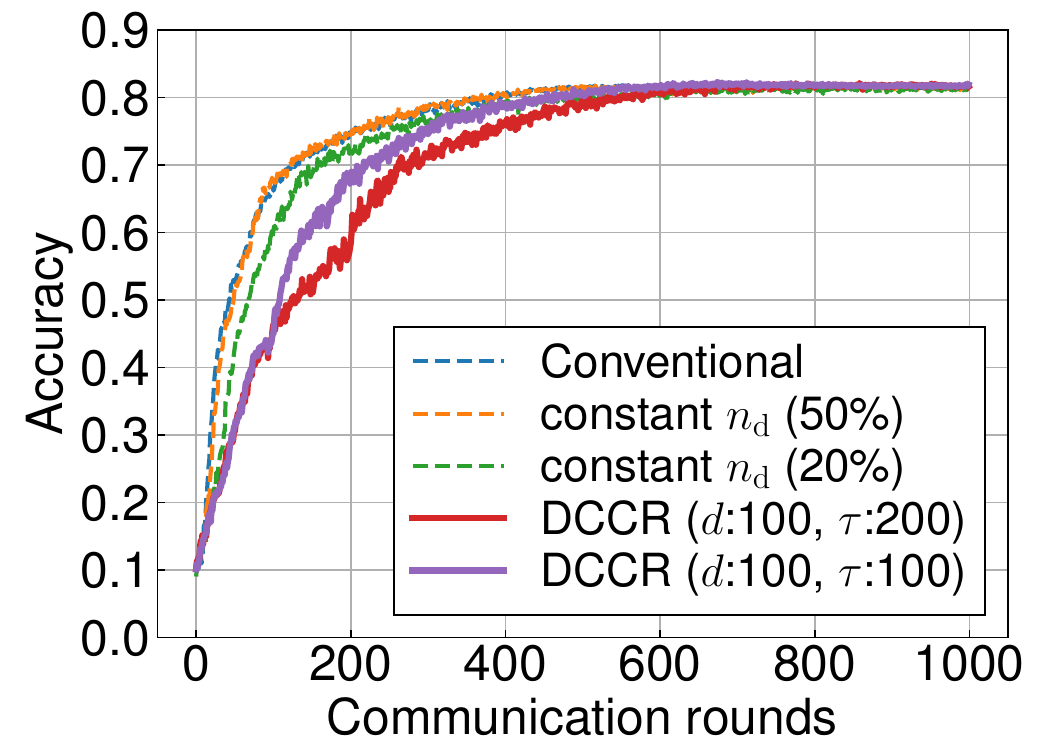}
  \caption{Prediction accuracy vs.\ communication rounds.}\label{fig:acc_r_fmnist}
\end{subfigure}
\begin{subfigure}[t]{\fsAcc}
  \centering
  \includegraphics[width=\textwidth]{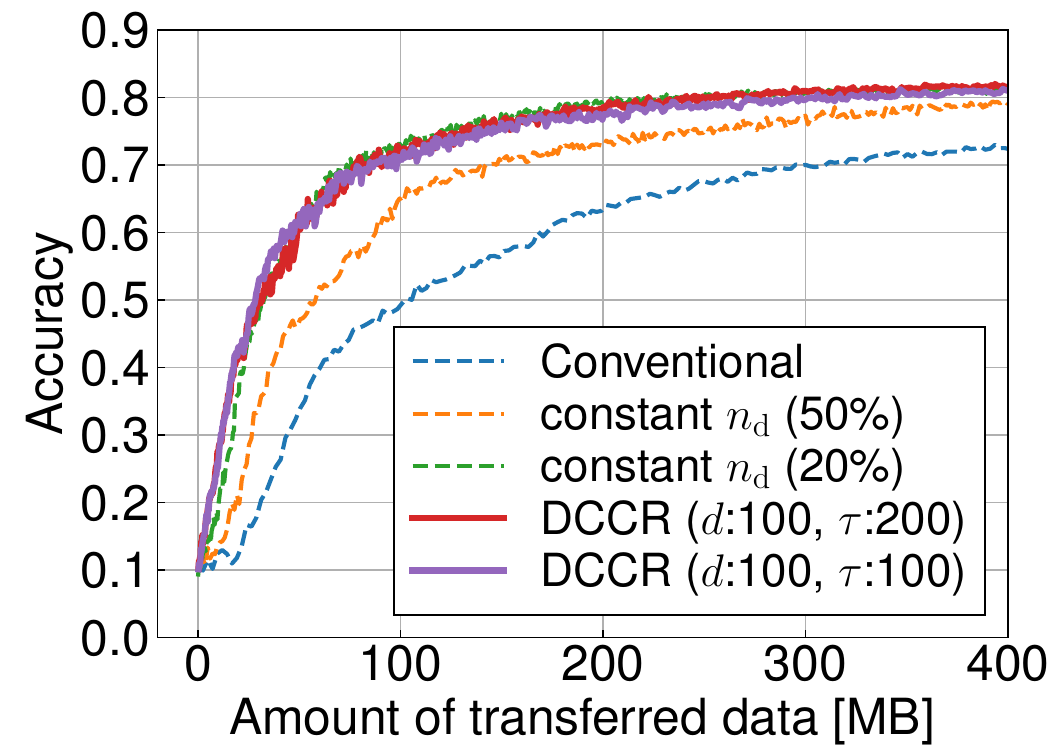}
  \caption{Prediction accuracy vs.\ communication costs.}\label{fig:acc_c_fmnist}
\end{subfigure}
\caption{Prediction accuracy as functions of communication rounds and communication costs when learning \gls{FMNIST}.
Limiting the data samples in distillation reduced the communication costs without degrading the prediction accuracy.
}\label{fig:acc_fmnist}
\end{figure}
\begin{figure}[!t]
\centering
\begin{subfigure}[t]{\fsAcc}
  \centering
  \includegraphics[width=\textwidth]{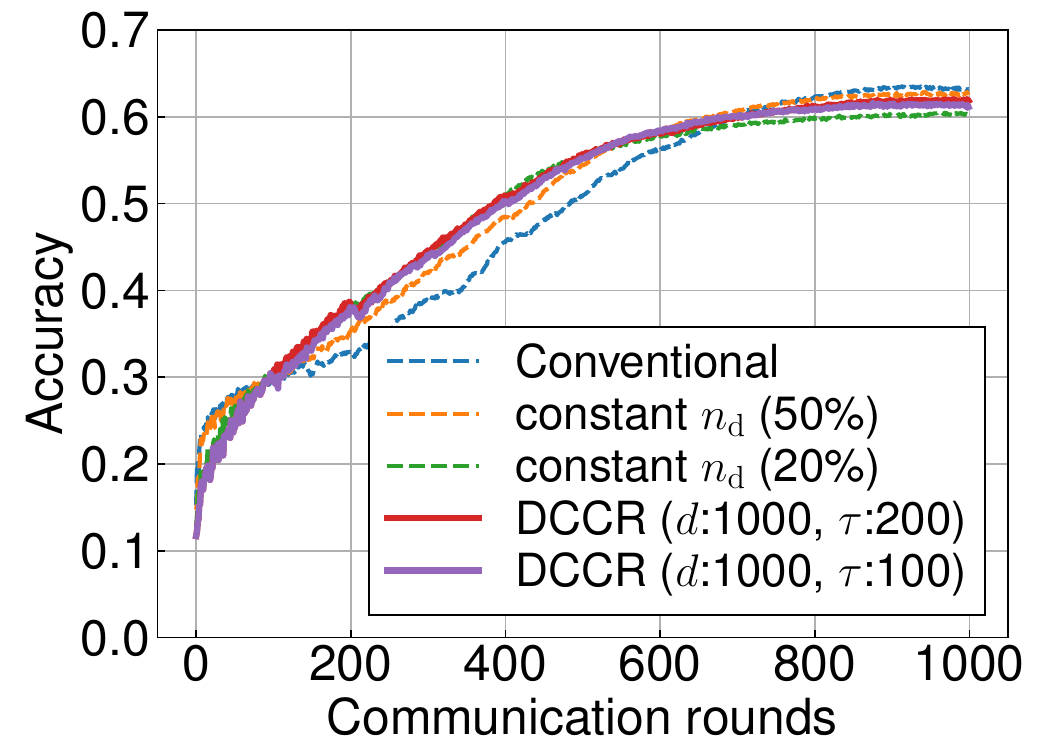}
  \caption{Prediction accuracy vs.\ communication rounds.}\label{fig:acc_r_cifar10}
\end{subfigure}
\begin{subfigure}[t]{\fsAcc}
  \centering
  \includegraphics[width=\textwidth]{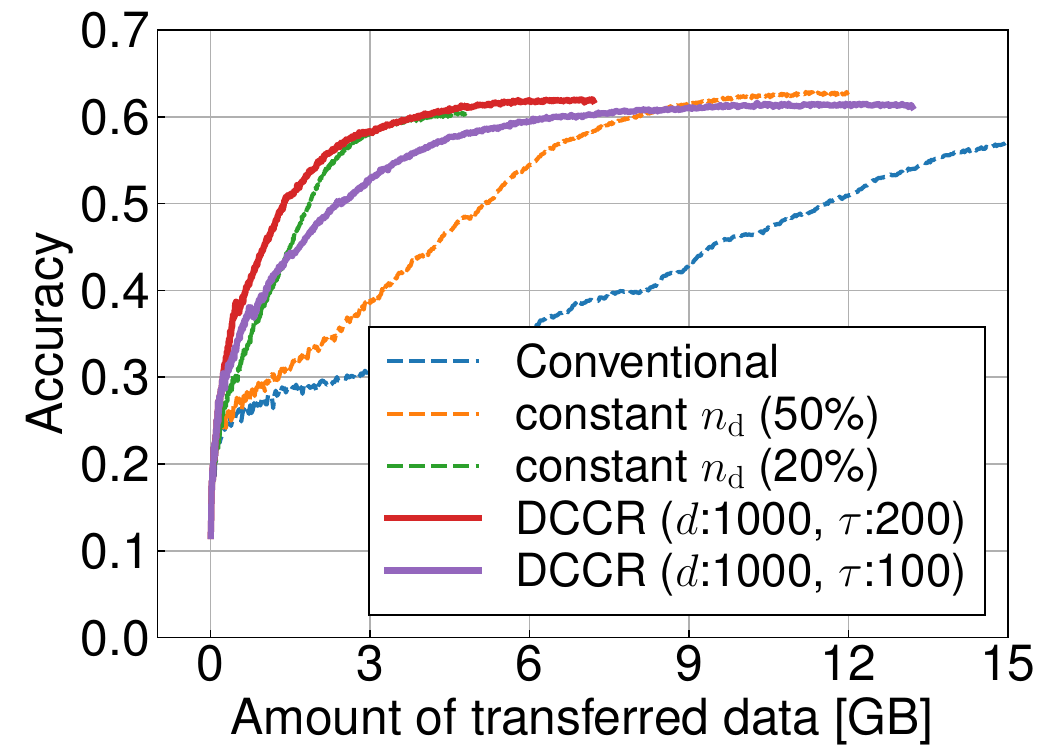}
  \caption{Prediction accuracy vs.\ communication costs.}\label{fig:acc_c_cifar10}
\end{subfigure}
\caption{Prediction accuracy as functions of communication rounds and the amount of transferred data when learning CIFAR10.
Compared with the conventional method, the \gls{DCCR} significantly reduced the total amount of transferred data.
}\label{fig:acc_cifar10}
\end{figure}

The results obtained for the \gls{FMNIST} and CIFAR10 datasets are shown in Figs.~\ref{fig:acc_fmnist}~and~\ref{fig:acc_cifar10}, respectively.
The left and right graphs in each figure show the prediction accuracy as a function of communication rounds and the amount of transferred data in the whole network, respectively.
% \begin{revhl}{1,4}
Figs.~\ref{fig:acc_fmnist}~and~\ref{fig:acc_cifar10} show the accuracy of \gls{CMFD} using different sizes $\NumBatchShared$ of the temporary dataset $\BatchShared$.
The dashed lines, labeled as ``Conventional,'' represent the accuracy of the original \gls{CMFD} that uses all data samples in $\sharedD$ for distillation.
Additionally, the figures illustrates the performance when either 20\% or 50\% of samples are randomly selected from $\sharedD$ during each communication round; these are labeled as ``constant $\NumBatchShared$.''
The figures also present the performance of \gls{DCCR} with two parameter settings.
Parameters $\NumDataTicks$ and $\TimeStep$ control the increase rate on $\NumBatchShared$ based on (\ref{eq:num_increase_rate}).
\Gls{DCCR} with $\tau=200$ uses less samples and is more communication efficient than that with $\tau=100$.
% \end{revhl}

\begin{figure}[!t]
\begin{subfigure}[t]{0.22\textwidth}
  \centering
  \includegraphics[height=10em]{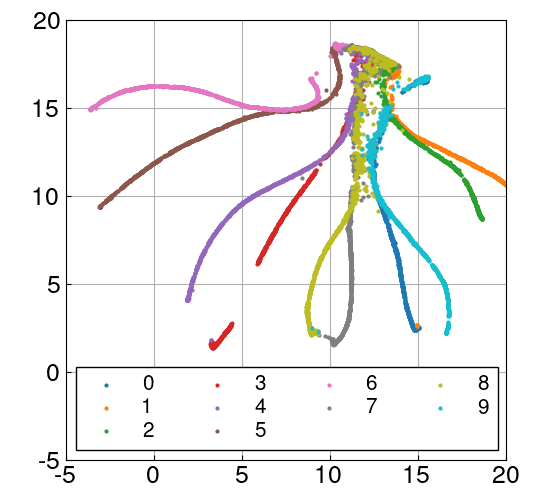}
  \caption{Color code:\\device ID}\label{fig:vis_fmnist_ada_d}
\end{subfigure}
\begin{subfigure}[t]{0.28\textwidth}
  \centering
  \includegraphics[height=10em]{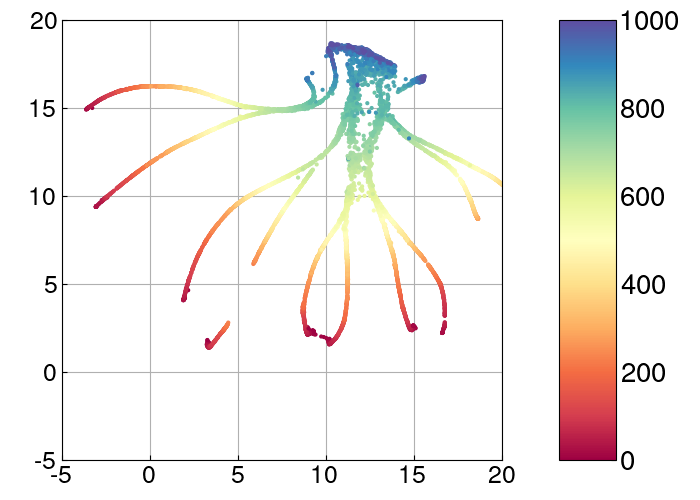}
  \caption{Color code:\\communication rounds}\label{fig:vis_fmnist_ada_r}
\end{subfigure}
\caption{Visualized learning process of \gls{FMNIST} using \gls{DCCR} with $\NumDataTicks=100, \TimeStep=200$.
Plotted points represent the two-dimensional projection of the local models during training.
In the left image, different colors represent different device models.
In the right image, points are colored by the communication rounds to track the progress of learning over time.
All local models converge to near positions at the end of the training.
}\label{fig:vis_fmnist_ada}
\end{figure}
\begin{figure}[!t]
\begin{subfigure}[t]{0.22\textwidth}
  \centering
  \includegraphics[height=10em]{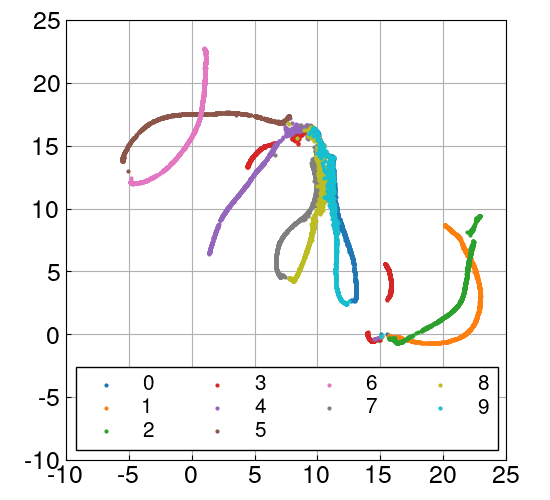}
  \caption{Color code:\\device ID}\label{fig:vis_fmnist_spr_d}
\end{subfigure}
\begin{subfigure}[t]{0.28\textwidth}
  \centering
  \includegraphics[height=10em]{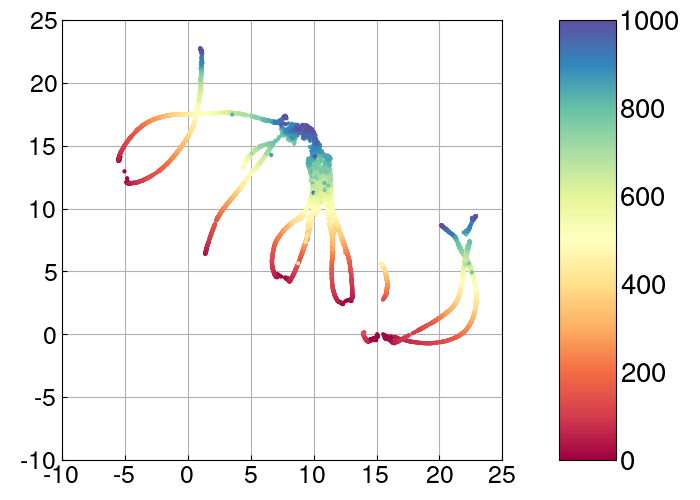}
  \caption{Color code:\\communication rounds}\label{fig:vis_fmnist_spr_r}
\end{subfigure}
\caption{Visualized learning process of \gls{FMNIST} when 20\% samples were used for distillation in each communication round.
Positions and colors represent the same information as Fig.~\ref{fig:vis_fmnist_ada}.
% The models of devices 1, 2, and 6 could not converge to the same position as the others.
Some models could not converge to the same position as the others.
}\label{fig:vis_fmnist_spr}
\end{figure}

% \begin{revhl}{4}
Although different strategies of $\NumBatchShared$ achieved similar accuracy in the end,
they differed in the amount of data required to reach that accuracy.
% \end{revhl}
As shown in Fig.~\ref{fig:acc_r_fmnist},
% \begin{revhl}{4}
\gls{CMFD} without \gls{DCCR} labeled as ``Conventional''
% \end{revhl}
achieved a higher accuracy than that with \gls{DCCR} at the beginning of the training
because it utilizes all samples of $\sharedD$ for distillation,
whereas distillation data are limited when using \gls{DCCR}.
However, \gls{DCCR} has advantages in terms of communication cost.
It is shown that \gls{DCCR} with $\NumDataTicks=100$ and $\TimeStep=100$ requires approximately 84\,MB data to achieve an accuracy of 70\% in Fig.~\ref{fig:acc_c_fmnist},
whereas the conventional scheme requires 300\,MB of data to achieve the same accuracy.
In addition, similar characteristics can be observed in Fig.~\ref{fig:acc_cifar10} showing the performance of training CIFAR10.
% 
% For comparison, we also evaluated the performance when the number $\NumBatchShared$ is constant.
% The results are shown in Figs.~\ref{fig:acc_fmnist}~and~\ref{fig:acc_cifar10}.
Although a slight difference can be observed in the accuracy
% \begin{revhl}{-}
with a constant $\NumBatchShared$ and \gls{DCCR},
% \end{revhl}
the local models did not converge with a constant $\NumBatchShared$ as discussed in the following subsection.

\subsection{Analysis with visualized function trajectories}
The visualized images of the convergence of the trained models $\Model_i$ when learning \gls{FMNIST} are presented in Figs.~\ref{fig:vis_fmnist_ada}~and~\ref{fig:vis_fmnist_spr}.
Fig.~\ref{fig:vis_fmnist_ada} shows the trajectory of the trained models when \gls{DCCR} ($\NumBatchShared=100$, $\TimeStep=200$) is applied. 
Fig.~\ref{fig:vis_fmnist_spr} also shows the trajectory with a constant $\NumBatchShared$, where 20\% samples were randomly selected in each round.
The colors of the plots in Figs.~\ref{fig:vis_fmnist_ada_d}~and~\ref{fig:vis_fmnist_ada_r} represent the device ID and number of communication rounds, respectively.
% \begin{revhl}{2}
The positions of the plotted points represent the \gls{UMAP}-based two-dimensional projection of the local models during training.
Since \gls{UMAP} is a distance-based dimensionality-reduction technique,
models that output similar values are projected to nearby positions.
When a line is drawn connecting the plots of each device's model,
the trajectory illustrates how the model changes gradually through gradient-descent-based training.
At the beginning of the training, local models are plotted at distant positions,
but they move closer to each other as training progresses.

Comparing the two schemes---\gls{DCCR} shown in Fig.~\ref{fig:vis_fmnist_ada} and a scheme with constant $\NumBatchShared$ shown in Fig.~\ref{fig:vis_fmnist_spr}---\gls{DCCR} achieved superior convergence.
Although both schemes required similar amounts of transferred data and attained comparable accuracy,
\gls{DCCR} outperformed in terms of convergence.
% \end{revhl}
% \begin{revhl}{5}
The underlying reason is that in the distillation steps with a small-sized $\NumBatchShared$, the models could not get close to each other sufficiently.
In contrast, \gls{DCCR} enabled the devices to utilize all samples in $\sharedD$ for distillation at the end of the training.
Consequently, \gls{DCCR} achieved better convergence compared with a scheme using constant $\NumBatchShared$.
% \end{revhl}

\begin{figure}[!t]
\begin{subfigure}[t]{0.22\textwidth}
  \centering
  \includegraphics[height=10em]{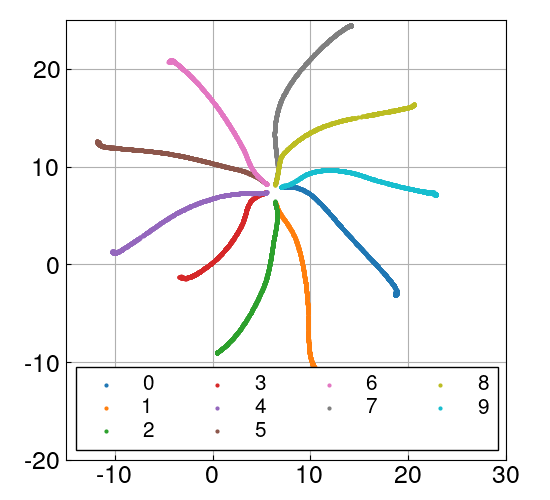}
  \caption{Color code:\\device ID}\label{fig:vis_cifar_ada_d}
\end{subfigure}
\begin{subfigure}[t]{0.28\textwidth}
  \centering
  \includegraphics[height=10em]{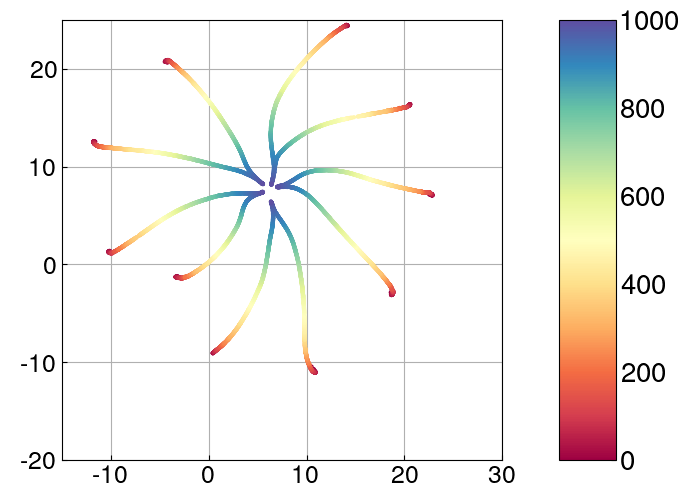}
  \caption{Color code:\\communication rounds}\label{fig:vis_cifar_ada_r}
\end{subfigure}
\caption{Visualized learning process of CIFAR10 using \gls{DCCR} with $\NumDataTicks=100, \TimeStep=100$.
Positions and colors represent the same information as Fig.~\ref{fig:vis_fmnist_ada}.
All local models successfully converged to the same positions at the end of the training.
}\label{fig:vis_cifar_ada}
\end{figure}

\begin{figure}[!t]
\nopagebreak
\begin{subfigure}[t]{0.22\textwidth}
  \centering
  \includegraphics[height=10em]{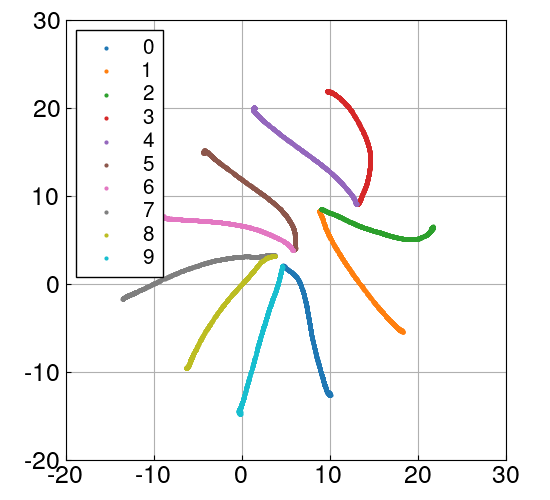}
  \caption{Color code:\\device ID}\label{fig:vis_cifar_spr_d}
\end{subfigure}
\begin{subfigure}[t]{0.28\textwidth}
  \centering
  \includegraphics[height=10em]{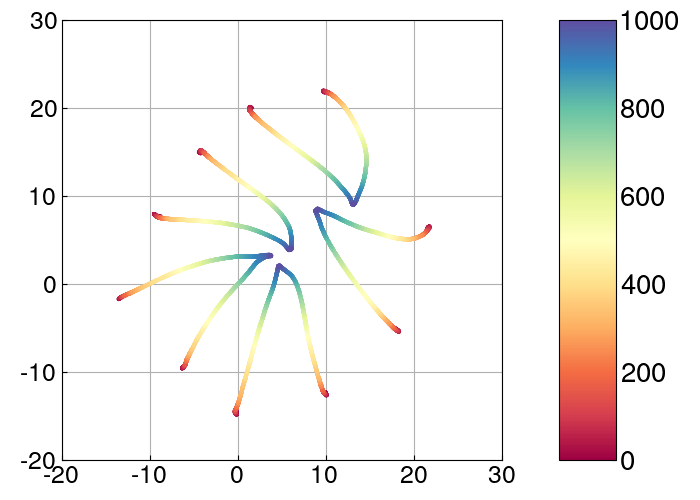}
  \caption{Color code:\\communication rounds}\label{fig:vis_cifar_spr_r}
\end{subfigure}
\caption{Visualized learning process of CIFAR10 when 50\% of samples were used for distillation in each communication round.
Positions and colors represent the same information as Fig.~\ref{fig:vis_fmnist_ada}.
The models could not converge to the same position.
}\label{fig:vis_cifar_spr}
\end{figure}

Figs.~\ref{fig:vis_cifar_ada}~and~\ref{fig:vis_cifar_spr} show the trajectories of the trained models when learnings the CIFAR10.
When we projected the prediction models onto a 2D space, we resampled 1000 samples from the shared data $\sharedD$
to reduce the computation time without degrading the visualization precision.
Furthermore, the characteristics similar to those of \gls{FMNIST} can be observed comparing Figs.~\ref{fig:vis_cifar_ada}~and~\ref{fig:vis_cifar_spr}.


\section{Conclusion}
We proposed a \gls{DCCR} that suppresses the transferred data in the \gls{DFD}.
With \gls{DCCR}, the devices first share a small number of distilled values with adjacent devices
and then gradually increase the number of shared values to improve the convergence performance.
We also developed a visualization scheme using dimensionality-reduction techniques.
The results showed the trajectory of how \gls{NN} models approach others in a function space.

% Our future direction includes \mytodo{future works}

\bibliographystyle{ieeetr}
\bibliography{main}

\begin{thebibliography}{10}

\bibitem{mcmahan2016communication}
B.~McMahan, E.~Moore, D.~Ramage, S.~Hampson, and B.~A. y~Arcas,
  ``Communication-efficient learning of deep networks from decentralized
  data,'' in {\em Proc.\ 20th Int. Conf. Artificial Intelligence and Statistics
  (AISTATS)}, (Fort Lauderdale, FL, USA), pp.~1273--1282, Apr. 2017.

\bibitem{kairouz2019advances}
P.~Kairouz {\em et~al.}, ``Advances and open problems in federated learning,''
  {\em Foundations and Trends{\textregistered} in Machine Learning}, vol.~14,
  pp.~1--210, June 2021.

\bibitem{uddin2019iot}
H.~Uddin, M.~Gibson, G.~A. Safdar, T.~Kalsoom, N.~Ramzan, M.~Ur-Rehman, and
  M.~A. Imran, ``{IoT} for {5G/B5G} applications in smart homes, smart cities,
  wearables and connected cars,'' in {\em 2019 IEEE 24th International Workshop
  on Computer Aided Modeling and Design of Communication Links and Networks
  (CAMAD)}, pp.~1--5, Sept. 2019.

\bibitem{beltran2022decentralized}
E.~T.~M. Beltr{\'a}n, M.~Q. P{\'e}rez, P.~M.~S. S{\'a}nchez, S.~L. Bernal,
  G.~Bovet, M.~G. P{\'e}rez, G.~M. P{\'e}rez, and A.~H. Celdr{\'a}n,
  ``Decentralized federated learning: Fundamentals, state-of-the-art,
  frameworks, trends, and challenges,'' {\em arXiv preprint cs.LG,
  arXiv:2211.08413}, Nov. 2022.

\bibitem{savazzi2020federated}
S.~Savazzi, M.~Nicoli, and V.~Rampa, ``Federated learning with cooperating
  devices: A consensus approach for massive {IoT} networks,'' {\em IEEE
  Internet Things J.}, vol.~7, pp.~4641--4654, May 2020.

\bibitem{elgabli2020gadmm}
A.~Elgabli, J.~Park, A.~S. Bedi, M.~Bennis, and V.~Aggarwal, ``{GADMM}: Fast
  and communication efficient framework for distributed machine learning,''
  {\em Journal of Machine Learning Research}, vol.~21, pp.~1--39, Mar. 2020.

\bibitem{taya2022decentralized}
A.~Taya, T.~Nishio, M.~Morikura, and K.~Yamamoto, ``Decentralized and
  model-free federated learning: Consensus-based distillation in function
  space,'' {\em IEEE Transactions on Signal and Information Processing over
  Networks}, vol.~8, pp.~799--814, Sept. 2022.

\bibitem{lalitha2019peer}
A.~Lalitha, O.~C. Kilinc, T.~Javidi, and F.~Koushanfar, ``Peer-to-peer
  federated learning on graphs,'' {\em arXiv preprint arXiv:1901.11173}, Jan.
  2019.

\bibitem{niwa2020edge}
K.~Niwa, N.~Harada, G.~Zhang, and W.~B. Kleijn, ``Edge-consensus learning: Deep
  learning on {P2P} networks with nonhomogeneous data,'' in {\em Proc.\ 26th
  {ACM} {SIGKDD} Int. Conf. Knowledge Discovery \& Data Mining}, (Virtual
  Conference), pp.~668--678, Aug. 2020.

\bibitem{anil2018large}
R.~Anil, G.~Pereyra, A.~Passos, R.~Ormandi, G.~E. Dahl, and G.~E. Hinton,
  ``Large scale distributed neural network training through online
  distillation,'' {\em arXiv preprint arXiv:1804.03235}, Apr. 2018.

\bibitem{zhang2018deep}
Y.~Zhang, T.~Xiang, T.~M. Hospedales, and H.~Lu, ``Deep mutual learning,'' in
  {\em Proc.\ 2018 IEEE/CVF Conf. Computer Vision and Pattern Recognition
  (CVPR)}, pp.~4320--4328, IEEE, 2018.

\bibitem{jiang2022model}
Y.~Jiang, S.~Wang, V.~Valls, B.~J. Ko, W.-H. Lee, K.~K. Leung, and
  L.~Tassiulas, ``Model pruning enables efficient federated learning on edge
  devices,'' {\em IEEE Transactions on Neural Networks and Learning Systems},
  pp.~1--13, 2022.

\bibitem{zhang2022fedduap}
H.~Zhang, J.~Liu, J.~Jia, Y.~Zhou, H.~Dai, and D.~Dou, ``{FedDUAP}: Federated
  learning with dynamic update and adaptive pruning using shared data on the
  server,'' {\em arXiv preprint arXiv:2204.11536}, 2022.

\bibitem{xu2021accelerating}
W.~Xu, W.~Fang, Y.~Ding, M.~Zou, and N.~Xiong, ``Accelerating federated
  learning for {IoT} in big data analytics with pruning, quantization and
  selective updating,'' {\em IEEE Access}, vol.~9, pp.~38457--38466, 2021.

\bibitem{nishio2019client}
T.~Nishio and R.~Yonetani, ``Client selection for federated learning with
  heterogeneous resources in mobile edge,'' in {\em Proc.\ 2019 IEEE
  international conference on communications (ICC)}, (Shanghai, China),
  pp.~1--7, May 2019.

\bibitem{sattler2020communication}
F.~Sattler, A.~Marban, R.~Rischke, and W.~Samek, ``Communication-efficient
  federated distillation,'' {\em arXiv preprint arXiv:2012.00632}, 2020.

\bibitem{uddin2020mutual}
M.~P. Uddin, Y.~Xiang, X.~Lu, J.~Yearwood, and L.~Gao, ``Mutual information
  driven federated learning,'' {\em IEEE Transactions on Parallel and
  Distributed Systems}, vol.~32, no.~7, pp.~1526--1538, 2021.

\bibitem{bishop2006pattern}
C.~M. Bishop and N.~M. Nasrabadi, {\em Pattern recognition and machine
  learning}, vol.~4.
\newblock Springer, 2006.

\bibitem{mcinnes2018umap}
L.~McInnes, J.~Healy, and J.~Melville, ``{UMAP}: Uniform manifold approximation
  and projection for dimension reduction,'' {\em arXiv preprint
  arXiv:1802.03426}, 2018.

\bibitem{maaten2008visualizing}
L.~van~der Maaten and G.~Hinton, ``Visualizing data using {t-SNE},'' {\em
  Journal of Machine Learning Research}, vol.~9, no.~11, pp.~2579--2605, 2008.

\bibitem{jeong2018communication}
E.~Jeong, S.~Oh, H.~Kim, J.~Park, M.~Bennis, and S.-L. Kim,
  ``Communication-efficient on-device machine learning: Federated distillation
  and augmentation under non-{IID} private data,'' {\em arXiv preprint
  arXiv:1811.11479}, Nov. 2018.

\bibitem{van2014renyi}
T.~Van~Erven and P.~Harremos, ``R{\'e}nyi divergence and {Kullback-Leibler}
  divergence,'' {\em IEEE Trans. Inf. Theory}, vol.~60, pp.~3797--3820, July
  2014.

\bibitem{xiao2017fashion}
H.~Xiao, K.~Rasul, and R.~Vollgraf, ``Fashion-{MNIST}: a novel image dataset
  for benchmarking machine learning algorithms,'' {\em arXiv preprint
  arXiv:1708.07747}, 2017.

\bibitem{krizhevsky2009learning}
A.~Krizhevsky and G.~Hinton, ``Learning multiple layers of features from tiny
  images,'' {\em Technical report, University of Tronto}, 2009.

\bibitem{Akiba2019-jz}
T.~Akiba, S.~Sano, T.~Yanase, T.~Ohta, and M.~Koyama, ``Optuna: A
  next-generation hyperparameter optimization framework,'' in {\em Proc. of the
  25th {ACM} {SIGKDD} International Conference on Knowledge Discovery \& Data
  Mining}, KDD '19, (Anchorage, AK, USA), pp.~2623--2631, Association for
  Computing Machinery, July 2019.

\end{thebibliography}

\end{document}